\documentstyle[11pt]{article}
\def\double{\baselineskip 20pt \lineskip 18pt} 
\def\be{\begin{equation}}
\def\ee{\end{equation}}
\def\bea{\begin{eqnarray}}
\def\eea{\end{eqnarray}}

\begin{document}
\begin{titlepage}

\begin{center}
\Large
{\bf The Embedding of Superstring Backgrounds in 
Einstein Gravity}

\vspace{.3in}

\normalsize

\large{James E. Lidsey$^{1}$}

\normalsize
\vspace{.6cm}

\vspace{0.6in}

{\em Astronomy Unit, \\ Queen Mary \& Westfield, \\Mile End Road, \\
LONDON E1 4NS, UK \\} 

\end{center}

\baselineskip=24pt

\vspace{3cm}

\begin{abstract}
\noindent
A theorem of differential geometry is employed to locally embed a wide
class of superstring backgrounds that admit a covariantly
constant null Killing vector field in eleven--dimensional, Ricci--flat
spaces. Included in this class are 
exact type IIB superstring backgrounds
with non--trivial Ramond--Ramond fields
and a class of supersymmetric string waves. The embedding spaces represent
exact solutions to eleven--dimensional, vacuum Einstein gravity. A
solution of eleven--dimensional
supergravity is also embedded in a twelve--dimensional, Ricci--flat
space.

\end{abstract}

\vspace{1in}

$^1$Electronic address: jel@maths.qmw.ac.uk

\end{titlepage}

\double 

Embedding theorems of differential geometry are important from both a
mathematical and physical point of view. They relate
higher-- and lower--dimensional theories of
gravity and give rise to classification schemes for different
spacetimes \cite{mac}. They can also lead to new solutions of
Einstein's field equations \cite{new}.  A well known theorem states
that any analytic, $n$--dimensional 
 Riemannian manifold can be locally
and isometrically embedded in a pseudo--Euclidean space of dimension
$n \le N \le n(n+1)/2$ \cite{eis}.  It is also known that the
pseudo--Euclidean space must have dimension $N \ge n+2$ if the
embedded space is Ricci--flat \cite{atleast}. 

On the other hand, there is a theorem due to Campbell that states that any
analytic, 
$n$--dimensional Riemannian manifold  can be locally 
embedded in a {\em Ricci--flat}, $(n+1)$--dimensional,
Riemannian space, where the extra dimension may be either space--like
or time--like \cite{campbell,ignore,rtz}.  Since the embedding space
is Ricci--flat, it represents an exact solution to the vacuum,
Einstein field equations in $(n+1)$ dimensions. One interesting
application of this theorem, therefore, is to generate new solutions to vacuum
general relativity by considering the embedding of
lower--dimensional solutions.

The embedding of four--dimensional electromagnetic and gravitational
plane waves in five dimensions was recently established by application
of Campbell's theorem \cite{apply}.  The purpose of the present paper
is to employ the theorem to embed exact ten--dimensional superstring
backgrounds in eleven--dimensional Einstein gravity. It is widely
thought that superstring theory represents a consistent quantum theory
of gravity and the study of classical solutions to the string
equations of motion that are exact to all orders in the inverse string
tension, $\alpha'$, is therefore important.

We consider spacetimes that admit a covariantly
constant null Killing vector field. 
The most general, $n$--dimensional spacetime admitting such a field
is the Brinkmann metric 
\cite{brinkmann}\footnote{Greek 
indices vary from $\mu = (0, 1, \ldots , n-1)$, 
lower case Latin indices run from
$i= (2, 3, \ldots , n-1)$ and upper case 
Latin indices take values in the range $A=(0,1, \ldots 
, n)$. The signature of spacetime is $( +, - ,
-, \ldots )$, a semicolon denotes covariant differentiation and
$\partial_{\mu}$ denotes 
partial differentiation with respect to $x^{\mu}$.}: 
\be
\label{generalmetric}
ds^2=2dudv +A_{\mu}(u,x^i)dx^{\mu}du -g_{ij}(u,x^i)dx^idx^j ,
\ee 
where
the Killing vector satisfies $l_{\mu ; \nu} =0$ and $l^{\mu}l_{\mu}=0$
and the light--cone coordinates $\{ u,v \}$ are defined by $l_{\mu}
\equiv \partial_{\mu} u$ and $l^{\mu} \partial_{\mu} v =1$,
respectively. The vector function, $A_{\mu}$, satisfies
$l^{\mu}A_{\mu} =0$ and has arbitrary dependence on $u$ and $x^i$.
The function $g_{ij}(u,x^i)$ is symmetric and positive
definite.  When the functional forms of the metric components satisfy
appropriate conditions, Eq. (\ref{generalmetric}) describes a wide
class of exact superstring backgrounds
\cite{exact,t,wave,F,K,exact1,exactreview,exactIIB}.  This includes
generalized string plane waves \cite{t,wave} and the $F$-- 
and $K$--models \cite{F,K}. (For a recent review of the different types of
known exact solutions see, e.g., Ref. \cite{exactreview}).

We begin by briefly reviewing Campbell's theorem and proceed to
establish the embedding of Brinkmann spacetimes.  We then apply the
embedding procedure directly to exact superstring backgrounds. In
particular, an embedding for a general class of type IIB backgrounds
with non--trivial Ramond-Ramond (RR) fields is found. Similar
embeddings are also determined for supersymmetric string waves
and solutions to eleven--dimensional supergravity.

In general  an analytic,  $n$--dimensional Riemannian manifold 
with metric ${^{(n)}}g_{\alpha\beta}
(x^{\mu})$ and line element 
${^{(n)}}ds^2 ={^{(n)}}g_{\alpha \beta} (x^{\mu} )
dx^{\alpha} dx^{\beta}$ 
may be locally embedded in a $(n+1)$--dimensional manifold with
a line element  
\be
\label{embed}
{^{(n+1)}}ds^2 =g_{\alpha\beta}(x^{\mu} ,\psi ) dx^{\alpha}
dx^{\beta} +\epsilon \varphi^2 (x^{\mu} , \psi) d\psi^2  , \qquad 
\epsilon =\pm 1 ,
\ee
if the metric coefficients 
$g_{\alpha\beta}= g_{\alpha\beta}(x^{\mu} ,\psi )$ and 
$\varphi = \varphi (x^{\mu} , \psi)$ 
satisfy certain restrictions. The problem 
of embedding the $n$--dimensional  manifold in 
$(n+1)$ dimensions is reduced to determining the appropriate 
functional forms for these components. 
This is achieved by introducing the set of functions 
$\Omega_{\alpha\beta} = \Omega_{\alpha\beta} (x^{\mu} , \psi )$
that satisfy the conditions 
\bea
\label{omega1}
\Omega_{\alpha\beta} = \Omega_{\beta\alpha} \\
\label{omega2}
{\Omega^{\alpha}}_{\beta ; \alpha} = \partial_{\beta} \Omega  \\
\label{omega3}
\Omega_{\alpha\beta}\Omega^{\alpha\beta} -\Omega^2 =-\epsilon 
{^{(n)}} R
\eea
on an arbitrary hypersurface $\psi =
\psi_0={\rm constant}$, where 
\be
{\Omega^{\alpha}}_{\beta} \equiv {^{(n)}}g^{\alpha\lambda}
\Omega_{\lambda\beta}, \qquad \Omega \equiv {^{(n)}}g^{\alpha\beta}
\Omega_{\alpha\beta}, \qquad {^{(n)}}R \equiv {^{(n)}}R_{\alpha\beta} 
{^{(n)}}g^{\alpha\beta}
\ee
and ${^{(n)}}R_{\alpha\beta}$ is the Ricci curvature tensor  
of the manifold calculated from ${^{(n)}}g_{\alpha\beta}$. We also 
require that  $g_{\alpha\beta}$ and $\Omega_{\alpha\beta}$ 
satisfy the coupled, partial differential equations
\bea
\label{metricdevelop}
\frac{\partial g_{\alpha\beta}}{\partial \psi} =-2\varphi
\Omega_{\alpha\beta} \\
\label{omegadevelop}
\frac{\partial {\Omega^{\alpha}}_{\beta}}{\partial \psi} =\varphi
\left( -\epsilon {^{(n)}}{R^{\alpha}}_{\beta}
+\Omega {\Omega^{\alpha}}_{\beta}
\right) +\epsilon g^{\alpha\lambda}  \varphi_{; \lambda\beta} 
\eea
and, moreover, that 
the components $g_{\alpha\beta}$ reduce to the $n$--dimensional metric
on the hypersurface $\psi =\psi_0$, i.e., that  
\be
\label{boundary}
g_{\alpha\beta}(x^{\mu} ,\psi_0 )= {^{(n)}}g_{\alpha\beta} (x^{\mu})  .
\ee 

It can then be shown that
Eqs. (\ref{omega1})--(\ref{omegadevelop}) correspond to 
the $(n+1)$--dimensional field equations of vacuum Einstein 
gravity,
\be
{^{(n+1)}}R_{AB} (x^{\mu} ,\psi_0 ) =0 ,
\ee
when Eqs. 
(\ref{metricdevelop}) and (\ref{omegadevelop}) are evaluated on 
the $\psi =\psi_0$ hypersurface \cite{rtz,Rippl}. However, if 
Eqs. (\ref{metricdevelop}) and (\ref{omegadevelop}) are 
valid in general, 
it can also be shown that  Eqs. (\ref{omega1})--(\ref{omega3}) are
satisfied for all  $\psi$  in the neighbourhood of $\psi_0$
\cite{campbell}. Thus, 
the Ricci tensor of the $(n+1)$--dimensional 
space vanishes for all $\psi$ in the neighbourhood of $\psi_0$,  
and since $\psi_0$ is arbitrary, 
this implies that the local embedding of the 
$n$--dimensional metric ${^{(n)}}g_{\alpha\beta}$ in a 
$(n+1)$--dimensional, Ricci--flat space is given by Eq. (\ref{embed}). 
Thus, 
the embedding is established by solving the set of 
equations (\ref{omega1})--(\ref{boundary}). 

We assume that $n=10$ and that 
the vector and tensor functions, $A_{\mu} (u,x^i)$ and $g_{ij}(u,x^i)$,
in Eq. (\ref{generalmetric}) are given by
\bea
\label{ansatzA}
A_u(u,x^i) \equiv \frac{1}{2} K(u, x^i) = \frac{1}{2} K_0(u, x^i) +
\frac{1}{2} K_{ij}(u) x^ix^j \nonumber \\
A_v \equiv 0 , \qquad 
A_i(u, x^j) \equiv -\frac{1}{2} M_{ij}(u)x^j
\eea
and 
\be
\label{ansatzg}
g_{ij}(u) \equiv f(u) \delta_{ij}  ,
\ee
respectively, 
where $\delta_{ij}$ 
denotes the eight--dimensional Kronecker delta and 
$M_{ij}(u) \equiv \partial_i A_j -\partial_j A_i$
represents an  effective field strength of the couplings $A_i$. The 
quantities $M_{ij}$, 
$K_{ij}$ and $f$ are arbitrary functions of $u$ only. The antisymmetric 
nature of $M_{ij}$ implies that $A_1$ is independent of $x^1$, etc. The 
function 
$K_0$ is an arbitrary solution to 
the eight--dimensional Laplace equation,
$\Delta K_0 =0$, where the Laplacian $\Delta \equiv \delta^{ij}\partial_i 
\partial_j$ is taken over the eight transverse directions. 
The only non--zero components of the Riemann tensor are 
${^{(10)}}R_{uiuj}$ and it 
follows that the one non--trivial component of the Ricci 
tensor is 
\be
\label{nonzero}
{^{(10)}}R_{uu} =-4f^{-1} \partial^2_uf +2 f^{-2} (\partial_u f)^2 
+\frac{1}{4} f^{-1} \Delta K +\frac{1}{16} f^{-2} M^2  ,
\ee
where $M^2 \equiv M_{ij}M_{ab} \delta^{ia} \delta^{jb}$. 
Moreover, since ${^{(10)}}g^{uu} =0$, the Ricci scalar curvature 
of the spacetime  (\ref{generalmetric}) vanishes, 
${^{(10)}}R=0$.

Given the ansatz (\ref{ansatzA}) and (\ref{ansatzg}), 
we now establish the local embedding 
of the general class of spacetimes (\ref{generalmetric}) in 
eleven--dimensional, vacuum Einstein gravity. The vanishing of the 
Ricci curvature scalar of the ten--dimensional metric implies that 
$\Omega_{\alpha\beta}=0$ is a 
consistent solution to Eqs. (\ref{omega1})--(\ref{omega3}). 
Eq. (\ref{metricdevelop}) then implies that
$\partial g_{\alpha\beta} /\partial \psi =0$. Thus,
the components $g_{\alpha\beta}$ of the eleven--dimensional 
embedding metric are independent of $\psi$
and Eq. (\ref{boundary}) is satisfied by choosing 
$g_{\alpha\beta}(x^{\mu})={^{(10)}}g_{\alpha\beta}(x^{\mu})$. 
The embedding is therefore completed once Eq. (\ref{omegadevelop})
has been solved and this equation represents 
the set of differential equations 
\be
\label{omegadevelop1}
{^{(10)}}g^{\alpha\lambda} \varphi_{;\lambda \beta} - 
{^{(10)}}{R^{\alpha}}_{\beta} \varphi =0  .
\ee
It is interesting that Eq. (\ref{omegadevelop1}) 
is independent of $\epsilon$, i.e., on whether the eleventh 
dimension is space--like or time--like.

All the components of ${^{(10)}}{R^{\alpha}}_{\beta}$ vanish 
apart from  ${^{(10)}}{R^v}_u = {^{(10)}}R_{uu}$. The 
form of $K(u, x^i)$ in Eq. (\ref{ansatzA}) is  such that  ${^{(10)}}{R^v}_u$ 
is a function of the light--cone coordinate $u$ only. 
This implies that it is consistent to assume that 
$\varphi =\varphi (u)$. It then follows that all the 
components of Eq. (\ref{omegadevelop1}) 
are trivially satisfied unless $(\alpha , \beta ) = 
(v,u)$. In this later case, 
Eq. (\ref{omegadevelop1}) simplifies 
to the second--order, ordinary differential equation 
\be
\label{varphiequation}
\left[ \frac{d^2}{d u^2} - {^{(10)}}R_{uu} (u) 
\right] \varphi (u)=0   .
\ee     
Thus, the embedding of the class of ten--dimensional spacetimes 
(\ref{generalmetric}), (\ref{ansatzA}) and (\ref{ansatzg})   is given by 
\be
\label{11}
{^{(11)}}ds^2 = 2dudv -\frac{1}{2} M_{ij}(u) x^j du dx^i 
+ \frac{1}{2}K(u, x^i) du^2 -
f(u) \delta_{ij} dx^idx^j +
\epsilon \varphi^2 (u) d\psi^2  ,
\ee
where $\varphi = \varphi (u)$ solves Eq. (\ref{varphiequation}). 
It may be verified by direct calculation of the eleven--dimensional 
Ricci curvature tensor that the manifold with metric (\ref{11})  
is indeed Ricci--flat when  Eq. (\ref{varphiequation}) is satisfied. 
We emphasize that the embedding is valid for {\em arbitrary} functions $\{ 
K_{ij}(u), M_{ij}(u), f(u) \}$. The function $K_0 (u ,x^i)$ is also 
arbitrary, subject to the condition $\Delta K_0 =0$. 

We now apply these results to the 
type II superstring theories. The 
zero--slope limit of the type IIA (IIB)  superstring is $N=2$, 
$n=10$ non--chiral (chiral) supergravity \cite{IIBzero,IIA}. 
The action for the 
Neveu--Schwarz/Neveu--Schwarz (NS--NS) sector 
of these theories in the string frame is \cite{bho}
\be
\label{NSNSaction}
S=\int d^{10} x \sqrt{|{^{(10)}}g|} e^{-\Phi} \left[ {^{(10)}}R +\left( 
\nabla \Phi \right)^2  - \frac{1}{3} \left( H^{(1)} \right)^2
\right] ,
\ee
where the graviton, ${^{(10)}}g_{\mu\nu}$, 
dilaton, $\Phi$, and antisymmetric two--form 
potential, $B^{(1)}_{\mu\nu}$, 
are the massless excitations. 
The field strength of the two--form is given by 
$H^{(1)}_{\mu\nu\lambda} \equiv \partial_{[\mu} B^{(1)}_{\nu\lambda]}$
and ${^{(10)}}g \equiv {\rm det}{^{(10)}}g_{\mu\nu}$. The 
background field equations derived 
from action (\ref{NSNSaction}) correspond to the conditions for the 
one--loop $\beta$--functions of the massless excitations to vanish. These 
may be solved by the {\em ansatz}  \cite{exact1,exactreview}
\bea
\label{background1}
B^{(1)}_{\mu\nu} =3B^{(1)}_{[\mu}l_{\nu ]} , \qquad 
B_v^{(1)} =0 , \qquad B_i^{(1)} (u, x^j ) =-\frac{1}{2} N_{ij}(u)x^j , 
\qquad 
\Phi =\Phi (u)
\eea
when the metric is 
given by Eqs. (\ref{generalmetric}), 
(\ref{ansatzA}) and (\ref{ansatzg}) with $f(u) =1$. The field 
strength 
$N_{ij}(u) \equiv \partial_i B_j - \partial_j B_i$ is a function of $u$ 
only. The $\beta_{uu}$--component  of the graviton $\beta$--function then 
takes the form 
\be
\label{Kij}
\Delta K +\frac{1}{4} M^2 - 
N^2 +4 \partial^2_u \Phi =0    ,
\ee
where $N^2 \equiv N_{ij}N_{ab} \delta^{ai}\delta^{bj}$ 
and this may be solved by choosing an appropriate 
functional form for ${\rm Tr}K_{ij}(u)$. All other components of the 
$\beta$--functions vanish identically. 
We refer to the class of superstring backgrounds 
that satisfy Eqs. (\ref{generalmetric}), 
(\ref{ansatzA}), (\ref{ansatzg}), 
(\ref{background1}) and (\ref{Kij}) as the 
`NS--NS backgrounds'. A geometrical argument may be employed to show 
that these backgrounds 
are exact to all orders in the inverse string tension \cite{wave,exactIIB}. 
The existence of the covariantly constant null Killing vector field 
implies that the Riemann curvature tensor is orthogonal 
to $l^{\mu}$. The same property 
is exhibited by $H^{(1)}_{\mu\nu\lambda}$ and the derivatives of the dilaton
field. This implies that all higher--order terms in the 
equations of motion vanish identically. 

We conclude immediately, therefore,  that the 
embedding of these exact NS--NS backgrounds in eleven--dimensional, 
Ricci--flat manifolds is given by Eqs. 
(\ref{varphiequation}) and (\ref{11}) and, in general, 
substitution of Eq. (\ref{Kij}) into Eq. (\ref{nonzero}) implies that 
\be
\label{generalnonzero}
{^{(10)}}R_{uu} (u) =\frac{1}{4} N^2(u) -\partial^2_u \Phi   .
\ee
As a specific example, we consider the case:
\be
\label{dilatonansatz}
N(u) \equiv {\rm constant}, \qquad 
\Phi (u) = a_0 + a_1 u + a_2 u^2 /2 ,
\ee
where $\{ a_i \}$ 
are arbitrary constants. This implies that ${^{(10)}} R_{uu}$ 
is itself a constant 
and the general solution to Eq. (\ref{varphiequation}) is therefore 
given by 
\be
\label{generalsolution} 
\varphi (u) = A 
\exp \left[ \sqrt{{^{(10)}}R_{uu}} u \right] +
B \exp \left[ -\sqrt{{^{(10)}}R_{uu}} u \right]  ,
\ee
where $A$ and $B$ 
are arbitrary constants. We emphasize that this embedding 
is valid for arbitrary functions $M_{ij}(u)$. 
In principle, a large class of non--trivial embeddings could be 
analytically calculated in this fashion 
for different functional forms of ${^{(10)}}R_{uu}$. 

The global ${\rm SL}(2,R)$ symmetry \cite{ht,2bsym} 
of the ten--dimensional 
type IIB superstring has recently been employed to generate 
new, exact type IIB backgrounds with non--trivial RR
fields from the above NS--NS backgrounds \cite{exactIIB}. 
These backgrounds are interesting because it is 
difficult to perturbatively 
calculate higher--order terms in the type IIB  theory 
due to the specific coupling of the RR sector \cite{RRalpha}. 
This sector 
consists of a pseudo--scalar axion field, $\chi$, a two--form 
potential, $B_{\mu\nu}^{(2)}$, and a four--form potential, 
$D_{\mu\nu\lambda\kappa}$ \cite{IIBzero,bho}. 
The complex scalar field $\lambda 
\equiv \chi +ie^{-\Phi /2}$ parametrizes the coset  ${\rm SL}(2,R)$/U(1) 
and transforms to $\tilde{\lambda} =(a \lambda +b)/(c \lambda +d )$ 
under a global ${\rm SL}(2,R)$ transformation, where $ad-bc =1$. 
The NS--NS and RR 
two--forms transform as a doublet, $\tilde{B}_{\mu\nu}^{(i)} = 
\left( \Theta^T \right)^{-1} B_{\mu\nu}^{(i)}$, where 
$B^{(i)}_{\mu\nu}$ represents a two--component vector and 
\be
\label{Theta}
\Theta \equiv  \left( \begin{array}{cc} 
a & b  \\ c  & d
\end{array} \right) 
\ee
is an ${\rm SL}(2,R)$ matrix. 
The four--form transforms as a singlet, $\tilde{D}_{\mu\nu\lambda\kappa} 
= D_{\mu\nu\lambda\kappa}$,  and the metric transforms to 
\be
{^{(10)}}\tilde{g}_{\mu\nu} = \exp \left[ ( \tilde{\Phi} -\Phi )/4 \right]
{^{(10)}}g_{\mu\nu}   .
\ee

The NS--NS backgrounds represent solutions to the 
type IIB string equations of motion with $\chi =B^{(2)}_{\mu\nu} = 
D_{\mu\nu\lambda\kappa} =0$. This implies that $\lambda =ie^{-\Phi /2}$
and applying a general 
${\rm SL}(2,R)$ transformation to these backgrounds generates a
new type IIB background \cite{exactIIB}:
\be
\label{newRR}
{^{(10)}}ds^2_{\rm IIB} =f(u) \left[ 2dudv +\frac{1}{2} K du^2 +A_i
dx^i du -\delta_{ij}dx^i dx^j \right]   ,
\ee
where 
\be 
\label{f}
f(u) \equiv \left[ d^2 +c^2e^{- \Phi (u)} \right]^{1/2}  .
\ee
This transformation generates non--trivial $\chi$ and $B^{(2)}_{\mu\nu}$ 
fields. Redefining the null variable 
$U \equiv \int^u du' f(u')$ and relabelling it as $u$ then 
implies that  the metric (\ref{newRR}) takes the form 
\be
\label{KRR}
{^{(10)}}ds^2_{\rm IIB} =2du dv 
+\frac{1}{2} \hat{K}(u, x^i)du^2 +A_i(u, x^i) dx^i du - f(u) 
\delta_{ij}dx^idx^j   ,
\ee
where 
\be
\label{Khat}
\hat{K} \equiv f^{-1}K   .
\ee

It is important that $\hat{K}$ is a quadratic 
function of the transverse coordinates  $x^i$, because in this case 
it can be shown that the type IIB 
backgrounds (\ref{KRR}) are 
exact to all orders in $\alpha'$ \cite{exactIIB}. 
Furthermore, the metric (\ref{KRR}) is precisely 
of the form given in Eq. (\ref{generalmetric}), 
where the metric components satisfy Eqs. (\ref{ansatzA}) 
and (\ref{ansatzg}) and $\hat{K}$ is identified 
with $K$. Thus, the only 
non--vanishing component of the 
Ricci--tensor of the spacetime 
(\ref{KRR}) is ${^{(10)}}R_{uu} = {^{(10)}}{R^v}_u$ and  
substitution of Eqs. (\ref{Kij}) and (\ref{Khat}) into Eq. (\ref{nonzero})
implies that 
\be
\label{nonzeroII}
{^{(10)}}R_{uu} (u) = 
-4f^{-1}\partial_u^2 f +2f^{-2} ( \partial_u f)^2 + 
\frac{1}{4} f^{-2} \left( N^2-4\partial^2_u \Phi \right)   .
\ee
Since this component is a function of $u$ only, we may conclude that an 
embedding  in an eleven--dimensional,  Ricci--flat 
space of the general class of type IIB backgrounds (\ref{KRR}) with 
excited RR fields is given by Eqs. (\ref{varphiequation}) 
and (\ref{11}).  

It is interesting to consider the background 
generated by the specific ${\rm SL}(2,R)$ transformation where 
$a=d=0$ and $c=-b=1$.  Since the RR axion is initially 
trivial $(\chi =0)$, this
transformation 
changes the sign of the dilaton field, $\tilde{\Phi} =-\Phi$, and 
inverts the string coupling, $\tilde{g}^2_s \equiv  e^{\tilde{\Phi}} = 
g_s^{-2} = e^{-\Phi}$. 
It therefore maps the strongly--coupled regime 
of the theory onto the weakly--coupled regime, and vice--versa. It 
also directly interchanges the NS--NS and RR two--forms and leaves 
$\chi =0$. Since the RR two--form 
is initially zero, the new background has a trivial 
NS--NS two--form and non--trivial RR two--form. 

Substituting Eq. (\ref{f}) 
into Eq. (\ref{nonzeroII}) then implies that  
\be
\label{Ricci}
{^{(10)}}R_{uu} = -\frac{1}{2} (\partial_u \Phi)^2 +2\partial^2_u
\Phi +\left( \frac{N^2}{4} -\partial^2_u \Phi \right) e^{\Phi}
\ee
and Eq. (\ref{varphiequation}) can now be solved exactly if 
Eq. (\ref{dilatonansatz}) is satisfied and the dilaton 
is a linear function of $u$ $(a_2 =0)$. 
If we further assume for simplicity that $a_0=0$, substitution  of 
Eq. (\ref{Ricci}) into Eq. (\ref{varphiequation}) implies that 
\be
\label{bessel}
\left[ \frac{d^2}{dz^2} +\frac{1}{z} \frac{d}{dz} 
+\frac{a_1^2}{2} z^{-2} - \frac{N^2}{4} z^{a_1-2}
\right] \varphi =0  ,
\ee
where $z \equiv e^u$. Modulo a constant of proportionality,
the general solution to Eq. (\ref{bessel}) is given by
\be
\label{besselsolution}
\varphi  = Z_{p} \left( Ne^{a_1u/2}/a_1 \right)  ,
\ee
where $Z_p$ represents a linear combination of 
modified Bessel functions 
of the first and second kind of order $ p = \pm 
\sqrt{2}i$. Identifying 
$Z$ with a modified function of the second  kind, for example,  
implies that $\varphi$ oscillates in the asymptotic 
limit $u \rightarrow -\infty$ and becomes exponentially damped 
when $u \rightarrow +\infty$ \cite{as}. This differs from the 
purely oscillatory or exponential behaviour 
of Eq. (\ref{generalsolution}) for the NS--NS background. 

The RR fields $\chi$ and $B_{\mu\nu}^{(2)}$ do not 
have an eleven--dimensional supergravity origin. 
However, Campbell's theorem implies that they may have a {\em 
geometrical} origin in the sense that all 
matter fields can in principle be geometrized in terms of higher 
dimensions. This is closely related to Wesson's interpretation 
of Kaluza--Klein gravity \cite{wesson,wesson1}. In this picture, it can 
be shown that five--dimensional, vacuum Einstein gravity 
gives rise to four--dimensional gravity with a {\em general} 
energy--momentum tensor  if one relaxes the condition  that 
physical quantities be independent  of the extra dimension, as was 
assumed in the original Kaluza--Klein theory. Thus, derivatives 
with respect to the fifth coordinate  are included and 
four--dimensional matter may be viewed as a manifestation of 
empty five--dimensional geometry. In effect, the 
geometry induces the matter. The extension of 
this interpretation to arbitrary dimensions has been considered in 
Ref. \cite{RRT}. It would be interesting to consider the 
relationship between the RR fields and eleven--dimensional 
geometry further in this context, although 
this is beyond the scope of the present 
paper. 

To conclude, we consider the class of NS--NS backgrounds:
\be
\label{SSWA} 
{^{(10)}}ds^2_{\rm SSW} =2dudv +A_u (x^i) 
du^2+A_i (x^i) dx^idu-\delta_{ij}dx^idx^j  ,
\ee
where 
the dilaton is fixed $(\Phi =0)$, the vector function 
$A_{\mu}$ is independent of $u$ and $v$ and the `chiral' 
constraint $A_i =-2B^{(1)}_i$ is imposed. 
This latter constraint implies that the vector--dependent terms in the 
graviton $\beta$--function (\ref{Kij}) cancel and this equation 
reduces to $\Delta A_u =0$. The background (\ref{SSWA}) 
may be embedded in the type IIA and type IIB 
superstring theories \cite{bho}. In the zero--slope limit, it
corresponds to a supersymmetric string wave
that admits eight constant Killing spinors and has precisely 
one--half of the spacetime supersymmetries unbroken \cite{exact1}. Thus, 
the embeddings that we have considered in this paper apply to this 
class of supersymmetric string waves. Indeed, since 
$M^2 =4N^2$, it follows from Eq. 
(\ref{nonzero}) that ${^{(10)}}R_{uu}= N^2/4={\rm constant}$ and 
the embedding metric is therefore given by
\be
\label{11SSWA}
{^{(11)}}ds^2_{\rm embed} = {^{(10)}}ds^2_{\rm SSW} -\varphi^2 d\psi^2  ,
\ee
where $\varphi$ is determined by Eq. (\ref{generalsolution}). 

The dimensional reduction of $N=1$, $n=11$ supergravity on a circle 
results in type IIA supergravity \cite{bho}. 
This feature has recently been employed 
to generate a new solution to the field equations of eleven--dimensional 
supergravity by lifting the ten--dimensional type IIA solution 
(\ref{SSWA}) to eleven dimensions \cite{bho}. 
The new solution generated in this fashion is 
\be
\label{super11}
{^{(11)}}ds^2_{\rm M} = {^{(10)}}ds^2_{\rm SSW} -dy^2   ,
\ee
where $y$ represents the coordinate of the eleventh dimension. 
The three--form antisymmetric potential is determined by the 
components $A_i(x^i)$. 
Eq. (\ref{super11}) generalizes the pp--wave solution found by 
Hull \cite{hullpp}. Since it 
admits a covariantly constant null Killing vector field, 
we may consider its embedding in a twelve--dimensional, 
Ricci--flat space along the lines discussed above. In particular,
the only non--trivial component 
of the Ricci--tensor ${^{(11)}}{R^{\alpha}}_{\beta}$ is 
${^{(11)}}{R^v}_u = 
{^{(10)}}R_{uu} = {\rm constant}$, where 
${^{(10)}}R_{uu}$ is calculated from the metric (\ref{SSWA}). 
This implies that we may immediately write down the 
embedding of the metric (\ref{super11}) in a 
twelve--dimensional, Ricci--flat space:
\be
\label{12}
{^{(12)}}ds^2 = {^{(10)}}ds^2_{\rm SSW} -dy^2-\varphi^2 d\psi^2   ,
\ee
where $\varphi$ is again given  by Eq. (\ref{generalsolution}). 

An important consequence of Campbell's theorem is that 
once the embedding of a $n$--dimensional manifold in a
$(n+1)$--dimensional, Ricci--flat space has been 
established, the procedure may be repeated 
indefinitely to embed the spacetime 
in Ricci--flat spaces of dimension $d \ge n+2$. Hence, 
the type IIA supersymmetric string wave (\ref{SSWA}) may 
be embedded in twelve--dimensional Einstein gravity by embedding 
the eleven--dimensional manifold (\ref{11SSWA}). Since this space 
is itself  Ricci--flat, Eqs. 
(\ref{omega1})--(\ref{omegadevelop}) are solved 
by $\Omega_{\alpha\beta} =0$
and $\varphi
 =1$. This implies that the twelve--dimensional embedding 
metric is given by ${^{(12)}}ds^2 = {^{(11)}}ds^2_{\rm embed} - dw^2$, 
where $w$ represents the coordinate of the twelfth dimension. 
However, this is formally 
identical to the metric (\ref{12}) if the extra dimensions 
are identified in an appropriate fashion. 
We remark that the twelfth dimension may be 
either space--like or time--like in both cases. Thus, 
the ten--dimensional type IIA 
supersymmetric string wave background and the solution (\ref{super11}) 
to eleven--dimensional supergravity may be locally embedded in the same 
twelve--dimensional, Ricci--flat space. 

In conclusion, therefore, we have employed Campbell's 
theorem to establish the local embedding 
of a general class of exact, ten--dimensional 
superstring backgrounds in eleven--dimensional,
Ricci--flat spaces. The embedded backgrounds admit 
a covariantly constant null Killing
vector field and the 
embedding spaces represent exact solutions of eleven--dimensional,
vacuum Einstein gravity. This is interesting 
because eleven--dimensional general relativity may be
directly related at a certain level to ten--dimensional superstring theory. 
There has been widespread interest
recently in the possibility that 
the five separate superstring
theories are related by discrete duality symmetries at a
non--perturbative level \cite{bho,ht,2bsym,duality,vafa,schM}.  This has
motivated the conjecture that they arise from a more fundamental
quantum theory (M--theory) \cite{schM,M}. 
Although the precise form of such a theory is at present
unknown, its low--energy effective action is eleven--dimensional
supergravity with a vacuum limit given by the Einstein--Hilbert action
\cite{m}.  In this sense, therefore, 
Campbell's theorem provides a
potential link between vacuum solutions 
of M--theory and ten--dimensional superstring backgrounds that
are valid in the strong curvature and 
strong coupling regimes. 

It would be of interest to establish similar 
embeddings for other classes of exact string solutions.  
In particular, we have considered backgrounds where the 
dilaton is a function of $u$ only, but it 
may also depend on the transverse coordinates $x^i$. When the 
metric admits a covariantly constant null Killing vector 
field and all 
massless excitations in the NS--NS sector 
are non--trivial functions of $x^i$ only, the 
backgrounds (\ref{generalmetric}) correspond 
to the class of `$K$--models' considered in Refs. \cite{K,exactreview}. 
There is also the related class of `$F$--models' that are characterized 
by two null Killing vectors 
and a chiral coupling between ${^{(10)}}g_{u \nu} $ 
and $B^{(1)}_{u \nu}$ \cite{F,K}. The simplest $F$--model 
is determined by a single function $F(x^i)$, where 
\be
{^{(10)}} ds^2 = F(x^i) du dv - \delta_{ij} dx^i dx^j, \qquad \Delta F^{-1} =0
\ee
and included in this class is 
the fundamental string solution \cite{DGHR}. Although Campbell's 
theorem implies that the embedding of these $F$--models is possible 
in principle, the 
form of the embedding will be more complicated than 
that considered 
in this work 
because the Ricci curvature scalar  of these backgrounds is in general 
non--zero. 

Finally, it is worth remarking that Campbell's theorem implies that
the type IIB backgrounds can also be embedded in twelve--dimensional,
Ricci--flat spaces. It would be interesting to explore possible
consequences of this feature within the context of the recently
proposed `F--theory' \cite{vafa}.  It has been conjectured that the
dimensional reduction of this twelve--dimensional theory on a
two--torus reproduces the type IIB theory, thereby providing a
geometrical interpretation of the ${\rm SL}(2, Z)$ symmetry of the
theory in ten dimensions. The interpretation of the D--instanton
background of the type IIB theory as the dimensional reduction of a
twelve--dimensional gravitational wave has recently been discussed by
Tseytlin \cite{tsey}. The question arises as to whether other 
ten--dimensional backgrounds
may be interpreted in a similar fashion by embedding them in twelve
dimensions.  This may provide further insight into the origin of the
RR fields $\chi$ and $B_{\mu\nu}^{(2)}$.

\vspace{.35cm}
{\bf Acknowledgments} The author is supported by the Particle 
Physics and Astronomy Research Council (PPARC), UK. We thank 
M. Piper for helpful discussions.

\end{document}